\documentclass{article} 
\usepackage{iclr2021_conference,times}

\usepackage{amsmath,amsfonts,bm}

\def\eqref#1{equation~\ref{#1}}

\def\1{\bm{1}}

\DeclareMathAlphabet{\mathsfit}{\encodingdefault}{\sfdefault}{m}{sl}
\SetMathAlphabet{\mathsfit}{bold}{\encodingdefault}{\sfdefault}{bx}{n}

\usepackage{hyperref}
\usepackage{url}
\usepackage{graphicx}
\usepackage{tabularx}

\newcommand\blfootnote[1]{%
  \begingroup
  \renewcommand\thefootnote{}\footnote{#1}%
  \addtocounter{footnote}{-1}%
  \endgroup
}

\title{Misspecification and Unreliable Interpretations in Psychology and Social Science}

\author{Matthew J. Vowels \\
UNIL\\
Lausanne, Switzerland.\\
\texttt{matthew.vowels@unil.ch} \\
}

\iclrfinalcopy 
\begin{document}

\maketitle

\begin{abstract}The replicability crisis has drawn attention to numerous weaknesses in psychology and social science research practice. In this work we focus on three issues that cannot be addressed with replication alone, and which deserve more attention: Functional misspecification, structural misspecification, and unreliable interpretation of results. We demonstrate a number of possible consequences via simulation, and provide recommendations for researchers to improve their research practice. Psychologists and social scientists should engage with these areas of analytical and statistical improvement, as they have the potential to seriously hinder scientific progress. Every research question and hypothesis may present its own unique challenges, and it is only through an awareness and understanding of varied statistical methods for predictive and causal modeling, that researchers will have the tools with which to appropriately address them.
\end{abstract}

\blfootnote{\copyright 2021, American Psychological Association. This paper is not the copy of record and may not exactly replicate the final, authoritative version of the article. Please do not copy or cite without authors' permission. The final article will be available, upon publication, via its DOI: 10.1037/met0000429}

\section*{Introduction}
Meta-researchers have increasingly drawn attention to the replicability crisis affecting psychology and social science \citep{aarts, Stevens2017, Marsman2017, Shrout2018}. A key element of the crisis relates to common and fundamentally problematic statistical practices, particularly those that affect the likelihood of false-positives such as statistical power. However, we would argue that there are a number of fundamental analytical issues which cannot be addressed through replication alone. These problematic practices relate to observational research and modeling in psychology and social science, and may be broadly categorized as issues with model misspecification, and unreliable interpretations. For clarity, we subdivide the issue of misspecification into functional and structural misspecification yielding three distinct issues.

 The first issue relates to the ubiquitous use of parametric, linear models, and a failure to consider more powerful techniques for both predictive and causal modeling. This issue is referred to as \textit{functional} misspecification. The second relates to another type of model misspecification, in this case, the use of \textit{structurally} misspecified implicit (\textit{e.g.}, multiple linear regression) or explicit (\textit{e.g.}, structural equation) causal models which do not sufficiently reflect the true structure in the data. The final issue relates both to how predictive models are often (mis)interpreted as causal models, and vice versa, and also to how these interpretations are likely to be unreliable given the models' underlying limitations, assumptions, and misspecification. All of these issues affect a researcher's ability to accurately model some aspect of the joint distribution of the data, for the purpose of predicting an outcome, estimating a causal effect, and drawing scientific conclusions. These issues share a particularly problematic trait, which is that they cannot be addressed or identified through replication alone. Indeed, in the presence of these issues, replications and larger sample sizes may serve only to reinforce flawed conclusions and therefore to hinder scientific progress.

 The goal of this paper is to introduce and draw attention to fundamental issues in common research practice through both didactic illustration and simulation, to provide some basic introductory theory where necessary, and to provide recommendations for improving practice. While some of these issues relating to research practice have been previously discussed (\textit{e.g.}, see \cite{Claesen2019, Scheel2020}), we believe it is extremely important to continue to encourage and stimulate consideration and engagement with the debate surrounding areas of possible analytical improvement. Furthermore, in spite of researchers having already made important recommendations for improving practice (\textit{e.g.}, \cite{Lakens2016, Scheel2020,Gigerenzer2018, Jostmann2016, Lakens2014, Orben2020}) we see relatively little change in the research communities of psychology and social science \citep{Claesen2019, Scheel2020}. 
 
 Following a review of the literature, the paper is split into three main parts. In Part 1, we describe how the typical models used in psychology are limited by their functional form and thereby functionally misspecified, and discuss the implications of this issue and possible ways to address it. Part 2 is concerned with structural misspecification in causal modeling, and how the typical models used in psychology and social science do not adequately reflect the true structure of the data. We discuss how this impacts interpretability, how a consideration for causal structure is essential when designing a model, and we identify some challenges associated with undertaking causal modeling. Part 3 introduces the notion of explainability as an alternative to interpretation, and as a means of deriving insight from predictive models. We discuss interpretation in light of the relevant points on limited functional form and structural misspecification covered in Parts 1 and 2, and consider how interpretations in psychology and social science tend to be a conflation of causal and predictive interpretations. Finally, we conclude this work with a discussion and by proposing four recommendations for improving practice.

\subsection*{Background - Existing Problems in Research}
A recent article titled `Declines in religiosity predict increases in violent crime - but not among countries with relatively high average IQ' was retracted from the Journal of Psychological Science on the basis of methodological weaknesses and political sensitivity. The Editor in Chief at the time, Steve Lindsay apologized on multiple grounds, and stated that ``In terms of science, Clark et al. may not be worse than some other articles published in Psych Science during my editorship...'' \citep{Lindsay2020}. This may suggest that methodological weakness, as described in terms of ``blurred distinctions between psychological constructs versus measures and speculations/extrapolations far removed from the data'' is somewhat par for the course in the ``young science'' \citep{Lindsay2020} of psychology. Indeed, over the last ten years, meta-researchers have drawn increasing attention to a purported crisis in the human sciences (particularly psychology) known as the replication crisis. The crisis has been discussed at length by many different meta-researchers (\textit{e.g.}, \cite{Oberauer2019, Botella2019, aarts, Stevens2017, Marsman2017, Shrout2018, Yarkoni2019}) who argue that research in the human sciences fails to replicate. 


The issues relating to replication are attributed to range of causes including a lack of understanding about and misuse of $p$-values and statistical tests \citep{Cassidy2019, Gigerenzer2018, Gigerenzer2004, Colquhoun2018, McShane2019}, misunderstandings about statistical power and low sample sizes \citep{Sassenberg2019, Baker2020, Correll2020}, pressure to publish \citep{Shrout2018}, academia and research being a strategy game with unscientific incentives \citep{Gigerenzer2018, dedeo2020}, a reluctance of journals to publish replications \citep{Martin2017, Gernsbacher2019}, and double-dipping and overfitting \citep{Fard2016, Kriegeskorte2009, Mayo2013, Yarkoni2017}.

Unfortunately, the problems do not end there. The range of identified methodological issues has broadened considerably to include aspects that stand independently of replication and which result in invalid inference. These issues relate to the ways researchers test theories \citep{Oberauer2019, Muthukrishna2019}; measurement problems \citep{Flake2020}; immature theories \citep{Scheel2020}; a lack of meta-analyses \citep{Schmidt2016}; a lack of assumptions testing \citep{ernst}; issues with the peer review process \citep{Heesen2020}; reporting errors \citep{Nuijten2016}; and a lack of research practice standardization \citep{Tong2019}. In particular, previous research has highlighted the tendency for the conflation of predictive and causal approaches \citep{Grosz2020, Yarkoni2017, Shmueli2010}, as well as overly generous claims and warped interpretations \citep{Yarkoni2019, Spellman2015,Scheel2020}. In our view, these latter two issues relate to some of the most fundamental aspects of research, and are discussed in this article. Choosing between a predictive or causal approach to research represents a critical decision that broadly impacts everything from the formulation of hypotheses to the conclusions drawn and interpretations made. 

 Altogether, there is evidence of a possible lack of understanding relating to the use of predictive models in psychology and social science, and we discuss this issue in detail in Part 1. If researchers choose to adopt a causal approach, it is equally important that they understand the associated pitfalls and challenges. For example, there is a relatively well established modeling technique known as Structural Equation Modeling (SEM) which explicitly encodes causal structure \citep{Kline2005,Blanca2018}. One point to note about the use of SEM in psychology and social science is that the way the technique is often presented and interpreted obfuscates its causal nature \citep{Grosz2020}. This leads to an awkward conflation of causal modeling with predictive interpretations, resulting in ambiguity and a lack of clarity regarding intentions and assumptions. It may be that researchers are unaware that their SEMs are explicitly causal and fail to sufficiently understand how the results from the analysis are underpinned by a number of restrictive (and often untestable) assumptions, which we discuss further in Part 2. 
 
If researchers choose prediction over causation, their scope for interpretation should be adapted accordingly. Meta-researchers have drawn attention to how it is common for psychology and social science researchers to use associational/predictive techniques to test otherwise causal hypotheses \citep{Shmueli2010, Yarkoni2017, glymour, Hernan2018, Grosz2020}. \citet{Shmueli2010} explains how ``the type of statistical models used for testing causal hypotheses in the social sciences are almost always association-based [\textit{i.e.}, predictive] models.'' Given this underlying interest in causal quantities, in Part 3 we present a means to derive useful insight relating to these quantities, even when a wholly predictive approach is used.

\section*{Part 1: Functional Misspecification}

In this part we address certain issues that may arise when using modeling techniques that have limited functional form. When we refer to the functional form of a model as being limited we mean that the model does not have the flexibility to sufficiently reflect the complexity of the relationships between variables, possibly resulting in poor predictive ability and biased results. This may also be referred to as functional misspecification. Identifying or deriving an adequately flexible functional form with which to model the relationship between variables, in circumstances where causal relationships are not of concern, is somewhat synonymous with the task of prediction. As such, the majority of this section will be written with consideration of its relevance to predictive modeling, where the goal is to learn a function that optimally maps predictor variables to outcome variables. However, a consideration for functional form is just as important for causal modeling, for which we may be tasked with embedding models representing the relationships between variables into a larger model representing the causal structure of the data generating process. For purposes of prediction alone, it suffices to be solely concerned with finding the optimal mapping function to achieve some desired level of predictive performance. Note that this goal is quite different from that of estimating a parameter of the distribution (\textit{e.g.}, a treatment effect). We expect models that reflect the structure of reality to also be robust predictors, but this is not necessarily the case the other way around; good predictive functions do not necessarily reflect the structure of reality.

We begin by briefly introducing some of the technical formalism behind predictive modeling, and list some of its wide ranging applications. Following this, we discuss the limitations of undertaking prediction using the two most common and basic methods used in psychology and social science: Correlation and linear regression models. We demonstrate how these methods, in the basic form adopted in psychology and social science, are fundamentally limited in their ability to adapt to and thereby model non-linearities present in the data. This motivates a need for more flexible, powerful, data-adaptive predictive methods. Previous research has highlighted that the use of such techniques is rare in psychology and social science, where it is much more usual to use models with restrictive linear functional form \citep{Yarkoni2017, Blanca2018}. Linear functions may be useful to consider for their computational efficiency and for their tendency to naturally \textit{under-fit} the data, thereby improving generalization particularly when the quantity of data is limited. However, these factors are not sufficient to fully explain the rarity of non-linear, powerful, and/or data adaptive techniques in psychology and social science, and we posit that a possible lack of awareness of these alternative methods is more likely.

 \subsection*{Applications and Basic Formalism}
 
The topic of identifying the optimal functional form with which to represent the relationship between variables is vast and well covered by many authors, particularly those in the field of machine learning in the context of prediction \citep{bishop, duda, murphy2}. Prediction has been described as ``the study of the association between variables or the identification of the variables which contribute to the prediction of another variable'' \citep{Blanca2018} and therefore relates closely to the more general task of identifying the optimal function that maps between sets of variables. The applications for predictive models are wide ranging, and include personalized medicine \citep{Rahbar2020}, time series forecasting \citep{Makridakis2020}, facial and object recognition \citep{alexnet, Jonsson2000}, and many others. Such techniques are therefore extremely valuable and influential in shaping our modern world.

 The basic formalism for predictive modeling is as follows: Researchers may be confronted with a dataset comprising samples from a population $(\mathbf{x}_i, \mathbf{y}_i) \in \mathcal{X} \times \mathcal{Y}$. In words, we have a set of samples of predictors or random variables\footnote{These variables are sometimes called `independent variables', but because they are usually \textit{not} independent, we avoid this potentially unhelpful terminology.}, which take on values in the set $\mathcal{X}$ and which are related to some outcome variables\footnote{These variables are sometimes called `dependent variables', but because other dependencies often exist we also avoid this terminology.} which take on values in the set $\mathcal{Y}$. If the outcome is binary or categorical, the task of prediction becomes equivalent to one of classification. The goal of prediction usually involves finding a mapping function $f:\mathcal{X}\rightarrow\mathcal{Y}$. We will use the terms \textit{predictive function} and \textit{predictive model} to refer to the mapping function used to make predictions.

\subsection*{The Common Assumption of Linear Functional Form}
 
\begin{figure}[!ht]
 \centering
\includegraphics[scale=0.5]{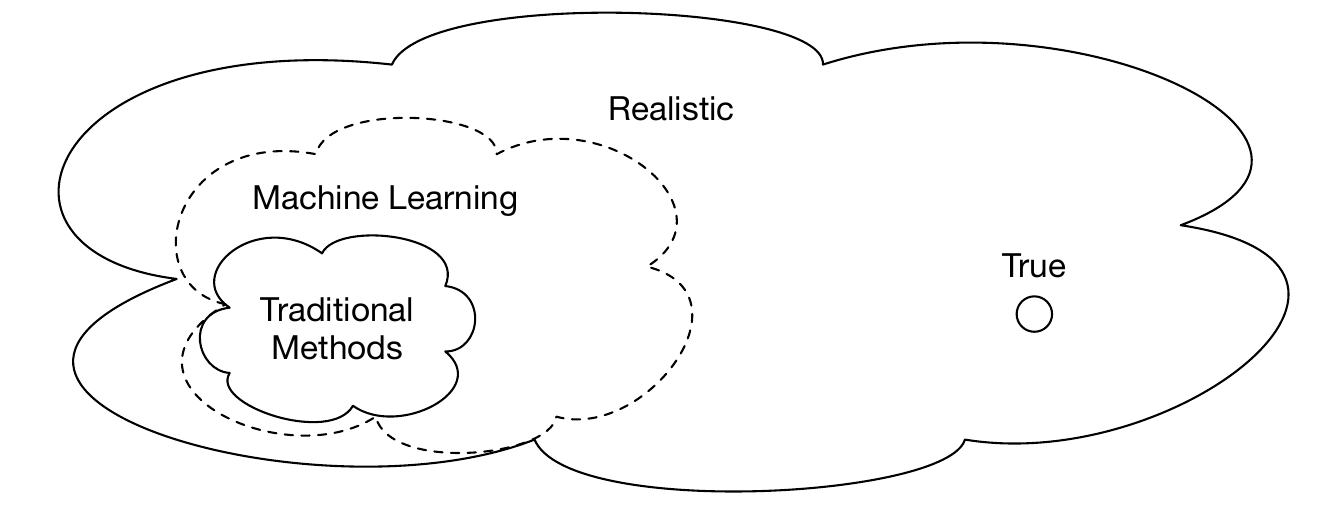}
\caption{ Traditional techniques such as linear regression may be severely limited in their capacity to model highly complex, non-linear data. Machine Learning methods may help to expand the coverage of realistic data distributions, but the true distribution may still lie outside. Combining flexible function approximation techniques from machine learning, with an incorporation of domain knowledge and model structure, can help us get as close as possible to modelling the true data distribution \citep{vanderLaan2011}.}
\label{fig:families}
\end{figure}

 Variations on simple measures of correlation and linear parametric models (including linear SEMs) were found to be the most frequently used modeling techniques in psychology research in recent years \citep{Blanca2018, Bolger2019}.\footnote{It might be argued that any arbitrary function can be represented as some linear sum of features, and that therefore all models are fundamentally linear. However, using such a broadly encompassing definition of the term `linear model' makes discussion pedantic. As such, we use the term to describe the typical linear regression model where the outcome is modeled as a linear sum of raw variables or low-order functions of these variables (such as exponents: $\mathbf{x}^1, \mathbf{x}^2$; and interactions: $ \mathbf{x}_1\mathbf{x}_2$ etc.).} The principal assumption associated with these models is that the true relationships between the variables are sufficiently represented as linear. Such models therefore have a limited functional form - they can only represent relationships between predictor and outcome variables that are able to be summarized in terms of a weighted sum. Of course, in reality the true relationships between variables may be highly complex and nonlinear. Indeed, assuming our dataset is sampled from a `true' population distribution, there exists an optimal functional form describing the functional relationships between the variables. Figure \ref{fig:families} illustrates how the set of traditional models provides limited overlap with the set of realistic data distributions, thereby having limited capacity to model complex real-world phenomena \citep{Coyle2020, vanderLaan2018, vanderLaan2014}. 
 
\begin{figure}[!ht]
  \centering
\includegraphics[scale=0.6]{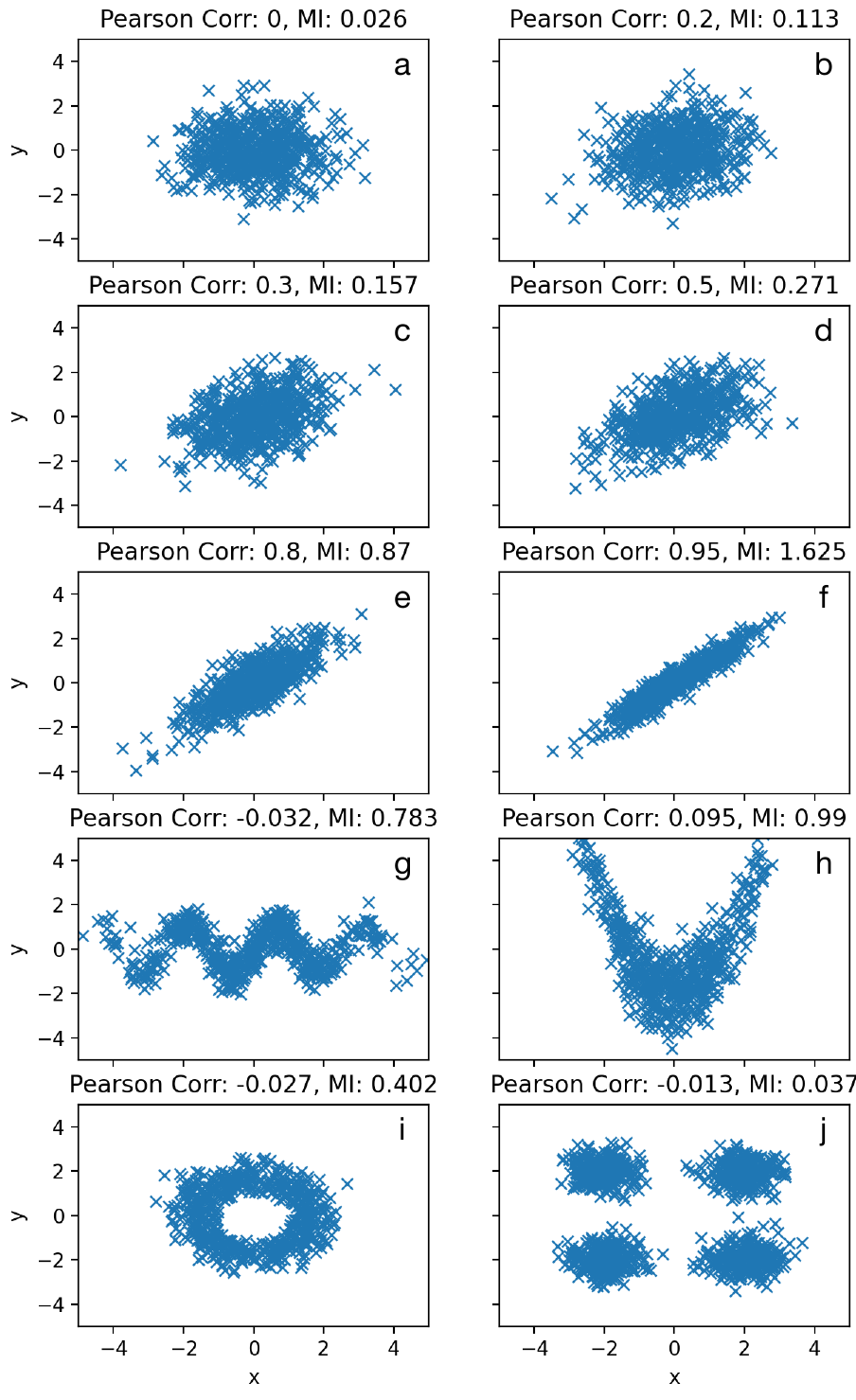}
 
\caption{ Simulations demonstrating the relationships between the Pearson measure of correlation, and the Mutual Information metric for measuring statistical dependence. The upper six plots depict linear bivariate relationships, whereas the lower four plots are non-linear.}
\label{fig:mi}
\end{figure}

 Correlation is often used as a measure for the association or statistical dependence between variables (\textit{i.e.}, to identify variables which may be good predictors). As one of the most common ways to measure dependence, there are two important aspects or issues relating to correlation to bear in mind. The first thing to note is that correlation is a non-linear metric for dependence. This can be observed by comparing the plots (a)-(f) in Figure \ref{fig:mi}. These plots depict a number of bivariate distributions along with their correlation coefficient (PCC). Compare, for instance, the change in the shape of the bivariate distributions between Fig. \ref{fig:mi}(a) and Fig. \ref{fig:mi}(b). This change corresponds with an increase in correlation of $0.2$ - there is not much to visually distinguish them. Now contrast this with a change of correlation from $0.8$ to $0.95$, which can be seen in Fig. \ref{fig:mi}(e) and Fig. \ref{fig:mi}(f) - this change is visually more striking, even though the change in correlation is actually smaller ($1.5$ rather than $0.2$). In other words, a change in correlation corresponds with a much greater increase in dependence when shifting at higher values (\textit{e.g.}, $0.7$ to $0.9$), than it does for lower values (\textit{e.g.}, $0.1$ to $0.3$). This first issue is important for researchers to understand when drawing conclusions about relative levels of correlation. Intuitively, a correlation of one (or, indeed, negative one) represents an asymptotic limit, and deviations in values of correlation far from this limit are less extreme than those close to it.
 
 The second, and perhaps more obvious aspect of note from Figures  \ref{fig:mi}(g)-(j) is that the PCC catastrophically fails to capture non-linear dependence. Here, sinusoidal, quadratic, or toroidal relationships (to take three examples) are associated with approximately zero correlation. This relates to the assumption of linearity: If the relationship between the two variables is linear, then correlation provides a measure of linear dependence; if the relationship is non-linear, then correlation may provide meaningless measures of dependence. In cases where the relationship is non-linear, researchers will need to either linearize the relationship (\textit{e.g.}, by creating a new variable which accounts for this non-linearity), or consider using an alternative measure of dependence.

One possible alternative to correlation, which serves as a more robust and general way to quantify statistical dependence, is Shannon Mutual Information (M.I.), which provides us with a measure for how much information one variable contains about another \citep{cover2006, Kraskov2004, Steeg2012, Steeg2013, Gao2003, Kinney2014}. M.I. is defined as:

\begin{equation}
    I(X;Y)=H(X) - H(X|Y)
    \label{eq:mi}
\end{equation} 

where $H$ is entropy. Entropy describes the degree of surprise or uncertainty associated with a distribution and is computed as $-\sum_n^N p(x_n) \log p(x_n)$ where $N$ is the number of datapoints in the sample distribution, $x_n$ is an individual datapoint and $p(x_n)$ is its corresponding probability.\footnote{Note that this expression is for the entropy of discrete distributions (\textit{e.g.}, Bernoulli). A similar expression exists for the entropy of continuous distributions and is known as the differential entropy.} Revisiting Equation \ref{eq:mi}, we therefore see that mutual information can be interpreted as the difference between the uncertainty in $X$ and the uncertainty in $X$ if we already know $Y$.  M.I. ranges between $[0, H(.)]$, such that when the two distributions are identical, $I(X;Y) = H(X) = H(Y)$. M.I. cannot be negative, and as such it is not able to indicate the `direction' of the association in the way that correlation can. However, this is an acceptable limitation given that many non-linear relationships are non-monotonic (\textit{i.e.}, they are not always either increasing or decreasing) and in such cases a notion of positive or negative direction is unhelpful. Overall, M.I. is quite different from correlation which, simply, indicates whether two variables with a linear functional relationship increase or decrease together. Correlation thereby represents only a loose proxy for shared information, whilst M.I. gives us an actual measure of statistical dependence between variables. The estimates for M.I. are also shown in Figure \ref{fig:mi}, and it can be seen that M.I. not only handles non-linear relationships between variables (compare the values of M.I. in plots (a)-(f)), but also increases linearly with the degree of dependence of the variables (plots (g)-(j)).

Linear regression is another very common modeling technique used for both predictive and causal modeling and constitute a relatively small sub-class within the class of Generalized Linear Models. There is one principal assumption for linear regression which is important for achieving both successful causal \textit{and} predictive modeling. Namely, that the outcome can be well approximated using a weighted linear sum of the input variables. Indeed, the linearity imposes a strong functional constraint that restricts the function's flexibility and is, therefore, an assumption about \textit{functional form} \citep{vanderLaan2011}. Linear methods are unlikely to match the functional form of realistic data distributions, and to get closer to the true functional form, researchers should consider using more flexible predictive methods.

In certain scenarios, collecting more data or undertaking a replication \textit{may} indirectly highlight problems relating to functional form. For instance, \citet{VowelsSpectralTutorial} provide an example demonstrating how the direction of a coefficient alternates depending on the start and end points of a longitudinal sample exhibiting sinusoidal/periodic fluctuation. However, in general, issues relating to insufficient functional form stand independently of sample-size and replication.

 \subsection*{Improving on the Functional Form of Linear Models}

In order to improve the predictive or associational performance of a predictive function, researchers may need to explore either \textit{feature engineering} approaches, or other functional approximation techniques such as those commonly used in machine learning. Introducing hierarchical structure within linear functions can improve the fit \citep{Yarkoni2019, Gelman2007, Bolger2019}, but even hierarchical linear models are constrained according to linear functional associations.

Feature engineering involves the substitution of raw input variables with functions of these raw variables, sometimes called \textit{features}. Depending on the functional form used to derive these features, the features themselves may then be linearly related to the outcome, facilitating better overall functional approximation. For instance, researchers may include more exotic basis functions such as sinusoidal functions \citep{VowelsSpectralTutorial} or kernels \citep{Scholkopf2019}, or simply combine features to form new ones (\textit{e.g.}, interaction features which are composed by multiplying two variables together). Feature engineering may thereby help to account for the non-linearities of the data in the features themselves, but in doing so, each feature may need to be carefully chosen or designed. For example, in Figure \ref{fig:mi}, the plot in the fourth row on the right has a simple basis function which is $\mathbf{x}^2$. While the raw values of $\mathbf{x}$ could not be used to model the outcome as part of a linear sum, the squared values could be used to essentially linearize the predictor in question. However, in real-world applications (\textit{i.e.}, research scenarios with real data) we will not know the functional form \textit{a priori} and it may be difficult to ascertain. For instance, the underlying function may not be an exact quadratic function $\mathbf{x}^2$, but some other, arbitrarily complex function. The feature engineering process may or may not be guided by knowledge about the domain of interest. For example, in the case of a time series with known seasonal variation (\textit{e.g.}, financial data exhibiting fluctuation due to the business cycle) the use of sinusoidal basis functions may be well justified and aid prediction and generalization \citep{hamilton}.

\begin{figure}[!ht]
 \centering
\includegraphics[scale=0.5]{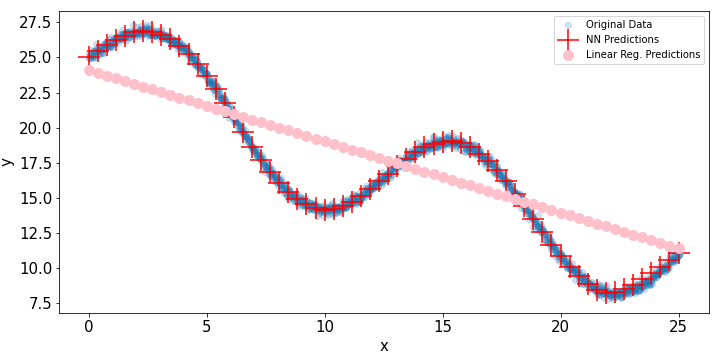}
 
\caption{ Demonstrates how linear functional forms cannot capture the non-linear relationships. In contrast, non-linear, data-adaptive techniques such as neural networks, can. }
\label{fig:NN}
\end{figure}

Besides generalized linear models with feature engineering, there exist many alternative and much more powerful function approximation techniques, such as those common in machine learning. These techniques are able to \textit{learn}, or estimate, functional relationships from the data themselves and can be used instead of, or in combination with, feature engineering. For instance, random forests \citep{breiman2001} comprise a group of decision trees that are capable of learning highly non-linear relationships and interactions between variables, without these interactions needing to be pre-specified. The mapping learned by the forest adapts to the data in order to minimize a performance objective (\textit{e.g.}, mean squared error). Neural networks are an alternative approach to function approximation which are also data-adaptive and are highly parameterized \citep{goodfellow}. They learn by iteratively updating their parameters according to an error signal until some criterion for convergence is met. An example of predictions from a simple neural network compared with those of a linear regressor on a bivariate problem is shown in Figure \ref{fig:NN}. It can be seen the neural network has fit the data almost perfectly, whilst the linear regression approximates the mean slope of the line, ignoring the cycling fluctuation. While prior knowledge may enable one to employ sinusoidal basis functions with linear regression in order to achieve a similar degree of fit, the advantage with neural networks is that such prior feature engineering is not required \citep{hornik2}.

\subsection*{Overfitting and Double-Dipping}

Overfitting and double-dipping refer to the consequences of various modeling practices which increase the fit of a model to a specific data sample, but which negatively impact the validity and generalizability of results. An awareness of overfitting becomes increasingly crucial when attempting to model non-linear functional relationships between variables. Overfitting and double-dipping have been extensively covered elsewhere, particularly in the machine learning literature (where overfitting is sometimes associated with what is known as the bias-variance trade-off) \citep{Belkin2019, bishop, murphy, Yarkoni2017, Mayo2013}. Prior research has highlighted how modeling practices that result in overfitting are common in psychology and social science, as well as a number of other fields, and have been noted for their possible contribution to the replicability crisis \citep{Shrout2018, Gelman2013, Yarkoni2017}. Even the common forward and backward method for variable inclusion constitutes data-driven overfitting practices which have the potential to significantly impact model generalizability and interpretability, and yet these practices are routinely included as part of standard statistical education and practice in psychology (\textit{e.g.}, see \citet{field}). When using powerful function approximation techniques, a consideration for overfitting is even more important. There are numerous techniques for mitigating issues with overfitting, including regularization, cross-validation (\textit{e.g.} k-fold or Leave One Out cross-validation), train-test splits etc. and it is important that researchers in psychology and social science familiarize themselves with these fundamental concepts, especially when modeling complex associations between variables.

\subsection*{Part 1 Summary and Discussion}
In Part 1, we described how models with limited functional form may be unable to represent the complex relationships between variables. Indeed, the assumption of linearity is often restrictive and has been previously noted to be problematic \citep{vanderLaan2011, Asuero2006, Daniel1999, Achen1977, King1986, Meehl1990, Taleb2019} and frequently ignored \citep{ernst}. Instances of functional misspecification may not be highlighted through replication, and this issue therefore stands independently. 

The typical analyses used in psychology and social science include simple measures of correlation, and various manifestations of linear regression. While such techniques are limited in their predictive capacity, there are many algorithms used in the field of machine learning which can learn an appropriately flexible functional form from the data themselves. When using more powerful techniques, it is especially important to validate models on an out-of-sample test set (\textit{e.g.}, by using a cross-validation method, or train/test splitting) in order to avoid overfitting. However, it is worth noting that overfitting (and the related problem of double-dipping) is also possible with simple linear models, and prior meta-research suggests that researchers may be unaware of these issues. 

Notwithstanding the argument for the use of flexible predictive models, we certainly do not wish to discount simple linear models completely. Indeed, in the limited sample regime it may be appropriate to approximate the relationships between variables using straight lines, or with low-order polynomials. However, one argument against this would be that a well-tested flexible model (well-tested in the sense that we are confident that it is not overfitting the sample) ought to arrive at a similar solution to a linear model, in cases where a linear relationship is optimal. Techniques such as k-fold cross-validation, or Leave One Out cross-validation, provide researchers with procedures to fit and evaluate flexible (or indeed linear) models on small samples. Furthermore, and as we will discuss in Part 2, linear models impose their own form of structural/causal bias, which becomes particularly problematic in situations exhibiting multivariate endogeneity.

Finally, the rarity of modeling techniques with powerful, data-adaptive functional form represents a possible missed opportunity in psychology and social science. We encourage researchers to consider the functional form of their models, and familiarize themselves with the associated pitfalls and limitations (\textit{e.g.}, overfitting), in order that they can get closer to modeling the true relationships underpinning the phenomenon under study. 

\newpage
\section*{Part 2: Causal/Structural Misspecification}

Prior research has highlighted a reluctance to adopt explicit causal approaches \citep{Grosz2020, Hernan2018}. Causal techniques provide the means to answer fundamental questions that help us to develop an understanding of the world \citep{Pearl2009, vanderLaan2011}. To the best of our knowledge, we are not aware of a well-established theory in psychology or social science which does not incorporate at least some level of consideration for cause and effect, and, if there is one, we would question its utility in so far as it can help us understand the world. Models which sufficiently align with the structure of reality may facilitate causal inference, even with observational (as opposed to experimental) data \citep{Glymour2001, Pearl2009, Pearl2016, Grosz2020} and have wide ranging applications including advertisement \citep{Bottou2013}, policy making \citep{Kreif2019}, the evaluation of evidence within legal frameworks \citep{Pearl2009, Siegerink2017}, and the development of medical treatments \citep{Petersen2017, vanderLaan2011}. There are a number of challenges associated with adopting a causal approach.

Structural misspecification represents one of the principal challenges associated with causal inference, and arises when the true causal structure and/or the functional form of the relationships between variables in the data generating process are not sufficiently reflected in a causal model. Structural misspecification results in biased effect size estimates which are not meaningfully interpretable. In this Part, we consider structural misspecfication in a restricted linear setting for purposes of demonstration. As we will show, even in this restricted setting, it is extremely important that the model sufficiently accounts for the true structure of the data in order that the resulting model is interpretable. This section is not intended as a technical guide to undertaking causal inference in general, but we nonetheless provide basic technical examples where relevant (for more information on causal inference see \textit{e.g.}, \cite{Pearl2009, Petersen2017, Pearl2016, Glymour2001, Angrist2001, Rubin2005, Gelman2007, Imbens2015}).

\subsection*{Recovering Causal Effects}
 Given the frequency with which psychologists and social scientists adopt linear regression methods to test causal theories \citep{Shmueli2010, Blanca2018}, it is extremely important that researchers understand the structural bias associated with the use of such models. In this section, we demonstrate how typical linear regression models used in psychology and social science impose a strong implicit causal/structural form which is unlikely to reflect the true causal structure of the data, even when the functional form is linear, and are therefore likely to be structurally misspecified. We show that, through a consideration of the causal structure of the phenomenon under study, one can nonetheless use linear regression to recover causal effects under a number of restrictive assumptions.

 \begin{figure}[!ht]
  \centering
\includegraphics[scale=0.4]{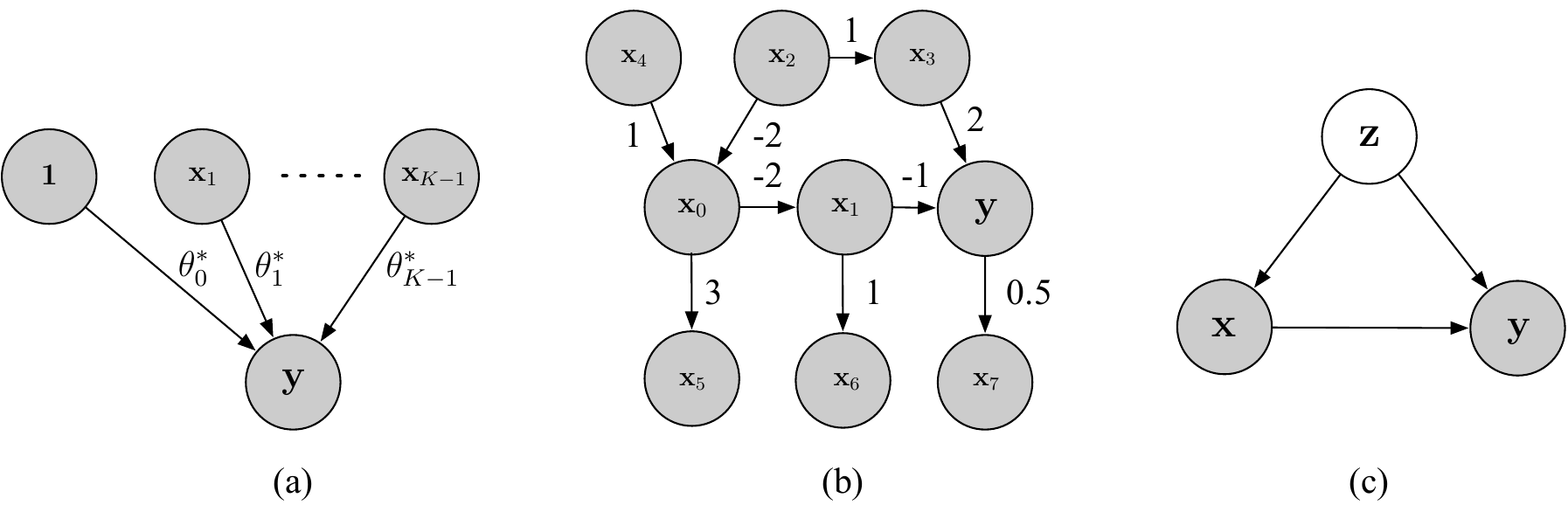}
\caption{ Example causal Directed Acyclic Graphs (c-DAGs). Example (a) depicts the case where all `predictor' or causal variables are exogenous (\textit{i.e.}, they have no causal parents and are independent of each other). This corresponds with the causal structure of a simple multiple regression, where the dependent outcome $\mathbf{y}$ is a linear sum of the $\mathbf{x}$ variables. The empirical causal effect of each variable is equivalent to the multiple regression coefficient estimates. Example (b) is adapted from \citet{Peters2017}. Example (c) depicts a graph with an unobserved confounding variable $\mathbf{z}$.}
\label{fig:simple}
\end{figure}

\subsubsection*{Multiple Regression Without Structural Misspecification}

In this section we demonstrate the strong, implicit structural form associated with multiple regression. We begin by demonstrating that multiple linear regression (in its basic form) is not misspecified (either functionally or structurally) with respect to the true data generating process when all predictors are exogenous (see structure in Figure \ref{fig:simple}(a)). In such a scenario, the resulting model is interpretable.

 If the true data generating process could be described as a weighted sum of a set of input variables, then our goal of prediction within the Ordinary Least Squares multiple linear regression framework would also be adequate for causal modeling, causal parameter estimation, or causal inference. Such a model might be depicted graphically as in Figure \ref{fig:simple}(a). In this scenario, there would exist parameters $\boldsymbol{\theta}^*$ (also known as effect sizes) which represent the true parameters, and our OLS-derived parameters would represent empirical/sample estimations thereof.

The graphs in Figure \ref{fig:simple} are known as causal Directed Acyclic Graphs (c-DAGs), and they represent a type of graphical representation often used in Structural Equation Modelling (SEM) \citep{Pearl2009, Koller2009, Rohrer2018}. The arrows indicate causal directional relationships between variables, parameterized by $\theta$, and the grey nodes indicate observed variables. The \textit{acyclicity} pertains to the restriction that there can be no closed loops (\textit{i.e.}, feedback) in the graph. Graph terminology (\textit{e.g.}, `parent', `ancestor', `descendant', `child') is useful in describing the top-level relationships between variables. For example, a node with an incoming arrow is a child of its parent variable, and further upstream or downstream variables are ancestors or descendants respectively. 

In general, the arrows in a c-DAG indicate causal dependencies, and there is no implied functional form that prescribes how the variables are combined at a node (\textit{i.e.}, there could be highly non-linear, adaptive functions with interactions). Furthermore, the nodes represent variables which may each be univariate or multivariate, and parametric or non-parametric. In other words, a node labelled $\mathbf{x}$ does not restrict the dimensionality or (non-)parameterization of $\mathbf{x}$ itself. For instance, a node $\mathbf{x}$ could comprise multiple predictors which do not conform to a parametric distribution. Hence, c-DAGs encode the fundamental essence of the causal structure, without imposing potentially irrelevant restrictions. We have included some extra information in the c-DAG of Figure \ref{fig:simple}(a) for the sake of demonstration. This particular c-DAG represents the intercept parameter of a multiple linear regression as a vector of ones multiplied by the parameter $\theta^*_0$. The structural equations for this graph may be represented in Equation \ref{eq:DAG1}:

\begin{equation}
    \begin{split}
        \mathbf{x}_{k=0} := \mathbf{1}\\
        \mathbf{x}_{k} := \mathbf{U}_{k}(0,1) \mbox{ for } k={1,...,(K-1)} \\
        \mathbf{y} := \sum_{k=0}^{K-1} \theta_k^* \mathbf{x}_k + \mathbf{U}_{y}(0,1) \mbox{ for } k={0,...,(K-1)} \\
    \end{split}
    \label{eq:DAG1}
\end{equation}

Let us assume that $\mathbf{U}_k$ and $\mathbf{U}_y$ are $N$-dimensional vectors of identically and independently distributed (i.i.d.) normally distributed random noise. The `$:=$' symbol denotes \textit{assignment} rather than equality. This distinction between assignment and equality is useful in reflecting the structural/causal direction of the arrows in the c-DAG. For example, the outcome $\mathbf{y}$ is a function of its inputs, and the equation should not be rearranged to imply that the inputs are a function of the outcome (the arrows point in one direction). Equation \ref{eq:DAG1} encodes that all the input variables are exogenous (\textit{i.e.}, completely independent of each other and determined only by i.i.d. noise) and that the outcome is determined by a weighted linear combination of these variables. In this setting we might understandably refer to the input variables as the independent variables, and the outcome as the dependent variable. As mentioned, these equations correspond with a simple multiple linear regression and can be solved to find $\boldsymbol{\theta}$ using OLS. We demonstrate this by undertaking a simulation for $K=4$ with $\theta^*_0 = 3.3$, $\theta^*_1 = 0.1$, $\theta^*_2= 0.3$ and $\theta^*_3 = 0.5$. We set $N=5000$ so that we do not have to be concerned about the stochastic variability associated with small samples, and the results are $\hat{\theta}_0 = 3.31$, $\hat{\theta}_1 =0.11$,  $\hat{\theta}_2 = 0.31$, and  $\hat{\theta}_3 =0.50$


From this demonstration it can be seen that the OLS regression successfully recovered $\hat{\boldsymbol{\theta}}$ close to $\boldsymbol{\theta}^*$. In this case, the data generating process directly matched the model we used to estimate the parameters and was therefore \textit{not} misspecified. When there is no misspecification, the estimated parameters may be interpreted as \textit{causal} parameters that tell us about the phenomenon (in this case, a simple, simulated phenomenon). Indeed, the parameters here can be interpreted as `one unit increase in $\mathbf{x}_1$ yields a $\theta_1$ increase in $\mathbf{y}$', as is common practice in psychology and social science. However, as we will see, the interpretability of this model is only possible because the structure of the data matched the structure of a multiple linear regression, equivalent to Figure \ref{fig:simple}(a), where all `predictors' are exogenous.

\subsubsection*{Multiple Regression - Structurally Misspecified for Realistic Phenomena}

In the previous section we showed how a simple multiple regression can be used to recover meaningful, causal parameter estimates, so long as the true causal structure of the data corresponds with the implicit causal structure implied by the multiple regression. There do exist other specific cases when a multiple linear regression can recover meaningful estimates, even when the true structure is more complex. Such cases arise when, for instance, the adjustment set (also referred to as the set of control variables) required to estimate a target quantity matches that reflected in the regression. However, the implicit causal structure of a linear regression is quite restrictive and, when modeling real-world data, it is likely to be \textit{both} functionally and structurally misspecified. In this section we demonstrate what happens when structural misspecification occurs.

Let us see what happens when we follow the same procedure to try to estimate some parameters for another simple data generating process which follows the example in Figure \ref{fig:simple}(b). We assume the following data generating structural equations (adapted from \cite{Peters2017}):

\begin{equation}
    \begin{split}
    \mathbf{x}_4 := \mathbf{U}_4, \; \; \; \; \mathbf{x}_2 := 0.8\mathbf{U}_2, \; \; \; \; \mathbf{x}_0 := \mathbf{x}_4 - 2\mathbf{x}_2 + 0.2\mathbf{U}_0, \; \; \; \;  \mathbf{x}_1 := -2\mathbf{x}_0 + 0.5\mathbf{U}_1, \\
    \mathbf{x}_3 := \mathbf{x}_2 + 0.1\mathbf{U}_3, \; \; \; \;  \mathbf{x}_5 := 3\mathbf{x}_0 + 0.8\mathbf{U}_5 , \; \; \; \;  \mathbf{x}_6 := \mathbf{x}_1 + 0.5\mathbf{U}_6, \\
    \mathbf{y} := 2\mathbf{x}_3 - \mathbf{x}_1 + 0.2\mathbf{U}_y, \; \; \; \;    \mathbf{x}_7 := 0.5\mathbf{y} + 0.1\mathbf{U}_7
    \end{split}
    \label{eq:DAG3}
\end{equation}

For these equations we have simplified the notation to make things clearer: $\mathbf{U}_k \sim \mathcal{N}(0,1)$. The structural process is still linear and the additive noise is Gaussian, so we do not yet need to worry about utilizing flexible function approximation techniques (such as those discussed in Part 1).  Note that $\mathbf{x}_3$ is only determined by $\mathbf{x}_2$, as well as its own exogeneous noise $\mathbf{U}_3$. This means that, if we perform surgery on these equations by, for example, setting $\mathbf{x}_3$ to a different value or distribution, we have cut off its dependence to its parent. Such graph surgery enables us to explore a range of causal queries such as interventions and counterfactuals, and is formalized by Pearl's \textit{do}-calculus \citep{Pearl2009}.

 Given the simple linear form in Equation \ref{eq:DAG3} for Figure \ref{fig:simple}(b), it is possible to traverse the paths in the c-DAG and to combine the effects multiplicatively. Such a process should be familiar to those who have studied path diagrams and SEM \citep{Kline2005}. For instance, the effect of $\mathbf{x}_0$ on $\mathbf{y}$ is the multiplication of the effect of $\mathbf{x}_0 \rightarrow \mathbf{x}_1$ with the effect of $\mathbf{x}_1 \rightarrow \mathbf{y}$. Together, we have the mediated path:  $\mathbf{x}_0 \rightarrow \mathbf{x}_1 \rightarrow \mathbf{y}$. According to Equation \ref{eq:DAG3} and Figure \ref{fig:simple}, the effect of $\mathbf{x}_0$ on $\mathbf{y}$ therefore corresponds with $-2 \times -1 = 2$. In this case, $\mathbf{x}_1$ is \textit{mediating} the effect of $\mathbf{x}_0$ on $\mathbf{y}$. Readers may already be aware of the issue relating to biased effect estimates which results from the inclusion of mediators in a regression analysis (see \textit{e.g.}, \cite{Cinelli2020, Rohrer2018, Pearl2009}). This issue is trivially demonstrated by comparing the regressions of $\mathbf{y}$ onto $\mathbf{x}_0$ whilst (a) adjusting for $\mathbf{x}_1$ and (b) and not adjusting for $\mathbf{x}_1$. Here, adjusting for a variable is equivalent to \textit{controlling} for it, but the adjustment terminology is more appropriate for structural scenarios \citep{Pearl2009}. First, the data are simulated according to Eq. \ref{eq:DAG3}, with $N=5000$. The bivariate correlations and $p$-values for each of these variables are shown in Table \ref{tab:pearsoncorrs}.

\begin{table}[h!]
\footnotesize
  \caption{Bivariate Pearson correlations and $p$-values for the DAG in Figure \ref{fig:simple}(b).}
  \label{tab:pearsoncorrs}
  \begin{tabular}{lrrrrrrrr}         \hline
 $r(p)$ & $\mathbf{x}_0$ & $\mathbf{x}_1$ & $\mathbf{x}_2$ & $\mathbf{x}_3$ & $\mathbf{x}_4$ & $\mathbf{x}_5$ & $\mathbf{x}_6$ & $\mathbf{x}_7$ \\ \hline
  $\mathbf{y}$ & .92(.00) & -.92(.00)& -.58(.00) & -.56(.00) & .76(.00) & .91(.00) & -.93(.00) & 1.00(.00) \\
  \hline
  \end{tabular}
\end{table}

The results in Table \ref{tab:pearsoncorrs} demonstrate a strong and statistically significant bivariate correlation between each predictor and the outcome. Now, when only using $\mathbf{x}_0$ as a predictor, we estimate the coefficient of $\mathbf{x}_0$ on $\mathbf{y}$ to be $\hat{\theta}_0 = 1.28$. Recall that the true effect of $\mathbf{x}_0$ on $\mathbf{y}$ is $2$. In spite of the large sample size, the output estimate is highly biased and does not seem to correspond with any of the causal parameters in the original simulation. Indeed, regardless of how large the sample size is, this coefficient estimate will converge to a value that is far from the true estimand. This is because the structure of the data generating process was not considered: We simply applied a linear regression to the data without accounting for the fact that the implicit structure of a linear regression does not match the structure in the data. In this situation, the multiple regression model might still have some (limited) utility as a purely \textit{predictive} function, but its parameters should not be interpreted as anything relevant to the causal structure of the phenomenon of interest because it is \textit{structurally misspecified}.

When confronted with the dilemma of multiple observed variables, typical practice in psychology and social science might involve using the forward or backward method for variable inclusion \citep{field}. Besides the problems associated with such practice (\textit{i.e.}, potential overfitting, as described in Part 1), including variables according to some predictive/associational heuristic is likely to result in structural misspecification because the rule for variable inclusion does not account for the underlying structure. Another approach might be to simply include all variables in the model. Indeed, all the $\mathbf{x}_k$ variables are highly and statistically significantly correlated with the outcome $\mathbf{y}$, so if we were not already aware of the implicit causal structure of linear regression, this might seem like a sensible thing to do. When we include all variables in the model, this results in $\hat{\theta}=-0.01$. Recall again that the true effect of $\mathbf{x}_0$ on $\mathbf{y}$ is $2$. The estimate of $-0.01$ is highly biased. This is because including all the variables in the model imposes the structure shown in Figure \ref{fig:simple}(a) which is clearly wrong.

Including $\mathbf{x}_0$ and the mediating variable $\mathbf{x}_1$ confirms that including mediating variables is problematic: The regression including both $\mathbf{x}_0$ and $\mathbf{x}_1$ yields $\hat{\theta} = -.94$. As expected, the effect of $\mathbf{x}_0$ on the outcome is highly biased, and of the opposite sign (\textit{i.e.}, negative rather than positive) to the true causal effect. It should now be clear that the use of what might be called naive multiple regression cannot yield meaningfully interpretable parameters unless the model corresponds with Figure \ref{fig:simple}(a) or  other exceptional cases, and this is unlikely for real-world phenomena. Indeed, it is arguable as to whether the interpretation of this parameter (and even its direction) is of any scientific value at all. 

\subsubsection*{Addressing Structural Misspecification Using Causal Inference Techniques}

We have seen that using naive multiple regression is inadequate when trying to estimate a causal effect from data with a non-trivial structure, even when the underlying functional form of the relationships is linear. Even where the structure is of relatively low complexity, the resulting coefficient estimates can be wildly biased. This illustrates that, regardless of whether the functional form matches the true functional form of the data (and in the linear simulations above, it did), it is impossible to recover meaningful effect size estimations with a misspecified model. In order to recover an unbiased estimate of the true effect, we need to understand techniques from the field of causal inference.

Structural Equation Modelling (SEM) is reported to be one of the most common methods used in psychology and social science \citep{Blanca2018}, and enables unbiased estimation of the parameters. The reliability of SEM depends on the structure of the model matching, or at least subsuming, the structure of the data generating process relevant to the estimation of the effect(s) of interest, as well as a number of additional restrictive assumptions \citep{Peters2017, VanderWeele}. The subsumption point relates to the notion that researchers, when faced with uncertainty about the structure of the data generating process, should choose to expand their model class rather than restrict it. In other words, it is a stronger assumption to impose the absence of a causal effect than to permit one. The choice to expand the model allows for the possibility of an effect in the data, whereas a removal of a causal link enforces an absence of dependency and thereby represents a strong model restriction that needs to be well justified before its imposition. Note that, in the presence of any such structural uncertainty, one clearly needs to suitably adjust the confidence of and qualify any associated interpretations and conclusions.

In practice, we rarely have access to the true model when we specify an SEM \citep{Damou2019,Wang2019b,Tenenbaum2002}. Indeed, as the SEM grows in complexity and/or its causal constraints, the chance of it becoming structurally misspecified increases. If certain assumptions are made, and we reduce our goal to the estimation of a specific and restricted set of effects (\textit{e.g.}, just the effect of $\mathbf{x}_0$ on $\mathbf{y}$), it may be sufficient to leverage domain knowledge and causal inference techniques to acquire a reliable estimate without having to correctly specify the full graph. For example, by identifying a sub-graph (sometimes referred to as a Markov blanket) which, for a given adjustment set, can be considered independently of the wider structure, one might significantly reduce the required complexity of the study design and analysis. Another way is by identifying what is known as a minimal backdoor adjustment set, and an example of this is presented in further detail below. In essence, one can identify the causal effect of $\mathbf{x}_0$ on $\mathbf{y}$ for the causal graph in Figure 4(b) (which involves nine variables in total) using only three variables ($\mathbf{x}_0$, $\mathbf{y}$, and $\mathbf{x}_3$ or $\mathbf{x}_3$). Many techniques for identifying specific effects exist and include the use of targeted learning \citep{vanderLaan2014}, g-computation \citep{Robins1986}, instrumental variables, propensity score matching, and regression discontinuity designs \citep{Blossfeld2009}. Such techniques have been extensively covered elsewhere \citep{Peters2017, Pearl2009, Imbens2015, Pearl2016, Angrist2001, vanderLaan2011, vanderLaan2014, Hernan2020}.

Backdoor adjustment involves identifying what are known as \textit{backdoor paths}, which are non-causal paths that affect the estimation of a target parameter or effect. An example of a backdoor path between $\mathbf{x}_0$ and $\mathbf{y}$ in Figure \ref{fig:simple}(b) is $\mathbf{x}_0 \leftarrow \mathbf{x}_2 \rightarrow \mathbf{x}_3 \rightarrow \mathbf{y}$. $\mathbf{x}_2$ and $\mathbf{x}_3$ are therefore part of what is known as the backdoor adjustment set; a set of variables which, if adjusted for, block the backdoor path. We can adjust for all the backdoor variables, or the minimal set sufficient to block the path (in our case, either $\mathbf{x}_2$ or $\mathbf{x}_3$ will do). Including $\mathbf{x}_0$ and $\mathbf{x}_3$ yields $\hat{\theta}= 2.00$.

We have now recovered an unbiased estimate of the effect of $\mathbf{x}_0$ on $\mathbf{y}$ (which was approximately equal to two), and we only needed to regress $\mathbf{y}$ onto two variables, despite our world knowledge dictating that at least eight (not including the outcome) were involved in the data generating processes as a whole (indeed, all variables in this simulation are highly and significantly correlated with the outcome). If we are also interested in the mediation through $\mathbf{x}_1$ then we can undertake separate regressions to break the problem down. The estimated parameters are then meaningfully interpretable insofar as they correspond with the parameters in the true data generating process. In other words, if $\theta = 2$, then every unit increase in $\mathbf{x}_0$ results in two units increase in $\mathbf{y}$.

\begin{figure}[!ht]
 \centering
\includegraphics[scale=0.5]{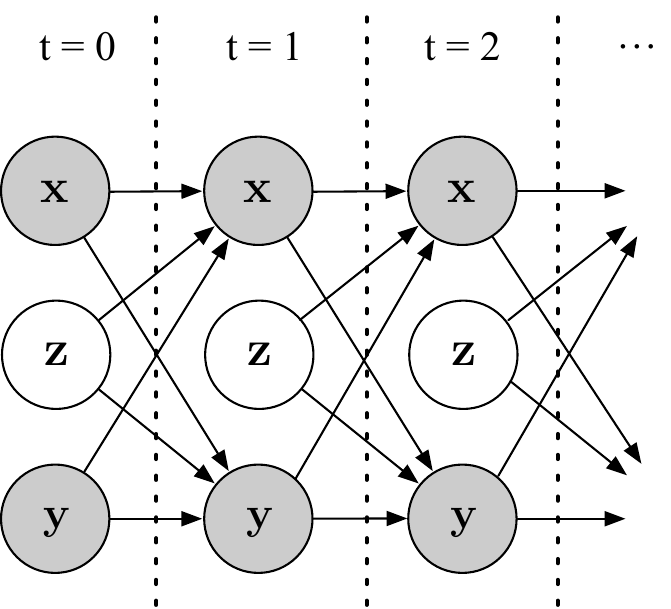}
 
\caption{ c-DAG for a time series setting, highlighting the complexity associated with identifying a particular causal effect, especially when there may be unobserved confounding \citep{Peters2017}.}
\label{fig:time}
\end{figure}

\subsubsection*{Longitudinal Challenges}
The constraint associated with the uni-directionality of time may be helpful in identifying the causal effect and correctly specifying a causal model. Indeed, time may represent a critical component of the data generating process. However, researchers must be aware of the additional analytical challenges associated with deriving estimates of causal effects for time-varying phenomena in order to properly take advantage of the benefits that longitudinal data can afford.

Figure \ref{fig:time} depicts a simple scenario with two variables, $\mathbf{x}$ and $\mathbf{y}$, and a hidden confounder $\mathbf{z}$. Each variable influences its own future as well as the future of the other variable. In the presence of the unobserved confounder the causal effect between $\mathbf{x}$ and $\mathbf{y}$ (however this might be defined) is \textit{unidentifiable}. That is, it cannot be derived or estimated from observational quantities. The complexity of this graph could grow further still if we include causal arrows between $\mathbf{x}$ and $\mathbf{y}$ (and potentially $\mathbf{z}$) for the same time point (\textit{i.e.}, $\mathbf{x}$ and time one influences $\mathbf{y}$ at time one), or if we add any additional (un)observed variables. Therefore, causal effects may be considerably difficult to identify in longitudinal data as a result of (1) difficulties avoiding the influence of confounders, and (2) difficulties distinguishing between intra- and inter-timepoint dependencies. The first point derives from the fact that conditions may change during the data collection window, and these changes manifest as unobserved variables that confound the quantities of interest. The second depends somewhat on the degree of temporal aggregation (\textit{i.e.}, are data being collected every hour, every minute, every second, or every millisecond?). 

These difficulties should not dissuade researchers from undertaking longitudinal research. On the contrary, longitudinal data provides additional information that cross-sectional data simply cannot. However, causal inference with time series data is challenging. In the context of causal inference with observational data, one may never be able to rule out the potential for unobserved confounding. This is particularly true when dealing with longitudinal measurements where there exist extended periods of time during which unobserved causes have opportunity to influence the phenomenon of interest. Interested researchers striving to make the most out of their longitudinal designs are directed to an accessible introduction of the topic, and its use in psychology, by \citet{Gische2020}. The application of causal inference to time varying phenomena is a very current and ongoing research topic in the fields of causality and machine learning \citep{Peters2017, Krishnan2016, Lohmann2012}.

\subsection*{Challenges, Assumptions, and Limitations of Causal Modeling}

Using only naive multiple linear regression models, we were unable to acquire a meaningful effect size estimate for non-trivial data generating process. Indeed, we used a relatively simplistic synthetic simulation to demonstrate that multiple linear regression yields meaningless estimates, but in real-world applications the graph may actually be significantly more complex which makes it extremely challenging to correctly specify the structure of the c-DAG, and therefore to use techniques such as backdoor adjustment. 

More generally, it is extremely difficult to obtain reliable effect size estimates from observational data concerning complex real-world social phenomena using these techniques. Indeed, \citet{Neal2020} provides a particularly extreme example where the naive approach yields as much as $407\%$ error in an effect size estimate. The infamous `crud' factor \citep{Orben2020, Meehl1990} that describes the way that ``everything [in social science] correlates to some extent with everything else'' also makes causal inference in social science and psychology particularly challenging.\footnote{See \citet{Orben2020} for a discussion of different definitions and origins of `crud'. We refer, in particular, to Lykken's  definition that considers crud to be the result of complex, multivariate interactions associated with psychological and social phenomena \citep{Lykken1968}.} This relates to the fact that social scientists are often concerned with the study of complex social systems with dynamic, multivariate interdependencies. Such systems may not exhibit readily identifiable cause and effect pairs \citep{Blossfeld2009}. As such, researchers are faced with the challenge of identifying suitable backdoor adjustment variables, as well as other structural entities such as colliders, mediators, instrumental variables, proxy variables etc. in order to facilitate the \textit{identification} of the causal effect using the observed data (for techniques, see \textit{e.g.}, \cite{Cinelli2020, Rubin2005, Imbens2015, Angrist2001, Pearl2009, Wang2019b, Damou2019}). 

In the same way that we chose to identify a \textit{single} causal effect using the backdoor adjustment method, it may be beneficial for researchers to attempt to simplify their causal research questions. For example, in contrast with the typical use of SEM in psychology and social science (where the researcher attempts to derive multiple effect estimates simultaneously), targeted learning adopts the philosophy by `targeting' a specific causal effect of interest, and orienting the analysis around its estimation using machine learning to reduce both functional and structural misspecification \citep{vanderLaan2011}. The `no free lunch theorem' familiar to machine learners applies here: causal inference yields the most information, but it may also be the most challenging \citep{Wolpert1997}. Attempting to undertake inference across multivariate, complex, linear SEM graphs is therefore extremely ambitious in light of its limited functional form and likely structural misspecification, and is highly unlikely to yield meaningful estimates.

Even once a researcher believes that they have accounted for the difficulties described above, and have simplified their research question or hypothesis, their consequent estimations then rest on various assumptions such as \textit{ignorability} - that there are no further latent/unobserved factors that have yet to be accounted for. Figure \ref{fig:simple}(c) depicts the presence of an unobserved confounder $\mathbf{z}$. Such an assumption may be strong, untestable, and unrealistic. Other assumptions may also be relevant, such as the positivity and stable unit treatment value assumptions (see \textit{e.g.} \cite{Imbens2015} for a discussion of these assumptions). Other problems relating to the restrictive nature of such assumptions, as well as the semantic leap required to move from a graph to real-world causal relationships, are discussed by \citet{Korb1997}, \citet{VanderWeele}, \citet{Dawid2008}, \citet{Kaiser2021}, and  \citet{Vowels2021DAGs}. It is important researchers familiarize themselves with all relevant assumptions and limitations before undertaking causal inference, and make them explicit in their work (\textit{e.g.}, when they use SEM).


The simulations above assumed linear structural equations. However, and as discussed earlier, c-DAGs do not restrict the functional forms relating the variables. Indeed, in real-world scenarios the assumption of linearity may impair the capacity of the model to estimate unbiased coefficients, in much the same way as it limited predictive models \citep{Coyle2020, vanderLaan2011, vanderLaan2014, vanderLaan2018}. The difficulties of effect estimation are therefore compounded by the difficulties associated with identifying an appropriate functional form for the dependencies between variables (\textit{i.e.}, identifying what \cite{Blossfeld2009} calls ``effect shapes''). Unless the structure of the model \textit{and} its functional form sufficiently match those of the true data generating process, \textit{and} we have an identifiable causal effect, the model may be misspecified and uninterpretable.

Notwithstanding these challenges, exploratory work can still be of value \citep{Shrout2018}. Part of the development process for SEMs (or, more generally, the underlying theory about the phenomenon) could involve causal directionality tests and validation via causal discovery techniques \citep{Peters2017, Vowels2021NSM, Vowels2021DAGs, Scholkopf2019, Glymour2019, Heinze2018, Spirtes2016}. Such techniques, at least under certain assumptions, may be able to test the directionality of the causal effects \citep{Goudet2019, Mooij2010}, identify backdoor adjustment set variables \citep{Gultchin2020}, estimate the magnitude of causal effects using flexible function approximation techniques \citep{Yoon2018, Shi2019}, induce the structure embedded in the data \citep{Glymour2019}, or infer hidden confounders from proxy variables using variational inference \citep{Louizos2017b}. However, these techniques are not without their own pitfalls and limitations, and, following a similar principle to \citet{Steegan2016}, researchers are encouraged to explore and test the dependencies of their exploratory insights and conclusions on their choice of analytical methodology.

\subsubsection*{Part 2 Summary}

 We described how difficult it is to obtain reliable causal effect size estimates, and we have also demonstrated how a failure to consider the causal structure may yield biased, meaningless effect sizes, regardless of whether the researcher adopts a predictive or causal approach, and regardless of sample-size or replicability. We provided one example of a causal inference technique known as backdoor adjustment, as a way to identify the causal effect of interest. Doing so enabled us to simplify the analytical problem from one of estimating all path coefficients in a complex graph, to one of estimating a specific effect by identifying variables from an adjustment set. In practice, identifying these backdoor variables is non-trivial, because it requires sufficient causal knowledge. Causal inference rests on a number of strong assumptions, perhaps the strongest of all being that of ignorability: That there are no unobserved confounders. Finally, researchers must also consider the functional form used to represent the causal dependencies between the variables. As such, problems with identifiability, ignorability, misspecification due to incorrect structure, and misspecification due to limited functional form have the potential to compound each other. In summary, it is important that researchers recognize the significant difficulties associated with estimating meaningful causal effects with observational data. However, it is worth also recognizing the value in persevering - as scientists we are primarily interested in answering causal questions. 
 

\section*{Part 3: Unreliable Interpretations}
In this part, we introduce explainability and interpretability, and describe how both structurally and functionally misspecified models may be neither explainable, nor interpretable. We discuss a range of problems relating to conflated and unreliable interpretations in psychology and social science. In our view, the conflation arises not just as a result of an alleged taboo against causal inference \citep{Grosz2020}, but also due to an apparent lack of understanding concerning the limitations associated with the interpretability of misspecified models with limited functional form and/or incorrect causal structure.

\subsection*{Explainability and Interpretability}

Scrutinizing the parameters of a model in a predictive sense is referred to as \textit{explaining}, in that we are explaining the behavior of the model, rather than \textit{interpreting} the model's parameters in relation to some external real-world causal phenomenon \citep{Rudin2019}. The explainabilty of a model describes the ease with which one can why a model makes a certain prediction or classification, based on its functional form or algorithmic rules \citep{Rudin2019}, and is therefore a term particularly relevant to predictive approaches. 

In contrast, we use the term interpretation to describe the process of using a model to understand something about the structure in the data or phenomenon, and the term is therefore of particular relevance to causal approaches. Explainability is not sufficient for interpretability because even if a model is structurally well specified, it cannot be interpreted if it is functionally misspecified. As we will show, linear models, such as those which are typical in psychology and social science, are not immune to problems affecting interpretability \textit{both} for reasons of functional misspecification as well as structural misspecification (see Parts 1 and 2).

\subsection*{Explaining Linear Models}

Linear models are deceptively simple to explain and interpret because the model coefficients are both directly interpretable as quantities in the real world (\textit{i.e.} causal effects), as well as useful in \textit{explaining} why the model makes a certain prediction (\textit{e.g.} a one unit increase in $\mathbf{x}_0$ corresponds with a $\hat{\theta_0}$ increase in $\hat{\textbf{y}}$). The deception occurs because of the two problems introduced in parts 1 and 2, notably functional and structural misspecification.

In the presence of functional misspecification, complex cancellation effects may render the coefficients of linear models incapable of explaining the model’s predictions \citep{Lundberg2020, Breiman2001b, Haufe2014}. The result will be that both the coefficients’ magnitudes (\textit{i.e.}, how \textit{much} the corresponding variable is associated with the outcome) as well as the ordinal levels of their importance (\textit{i.e.} which ones are most associated with model outcome) are arbitrary.\footnote{This applies regardless of whether the coefficients are standardized.} In order to demonstrate the issue of (un)explainability in the presence of misspecified functional form, we generate a synthetic example, closely following that of \citet{Lundberg2020}.\footnote{Full code for the original example can be found here: \url{https://github.com/suinleelab/treeexplainer-study/}.} The relationship between the outcome and two particular features in a semi-synthetic dataset is modified to include an increasing amount of non-linearity following the relationships in Equation \ref{eq:linprobs}.

\begin{equation}
    \mathbf{y} = \sigma((1-q)(0.388\mathbf{x}_1 - 0.325) + q(1.714\mathbf{x}_1^2 - 1) +  1.265\mathbf{x}_2 + 0.0233)
    \label{eq:linprobs}
\end{equation}

\begin{figure}[!ht]
 \centering
\includegraphics[scale=0.55]{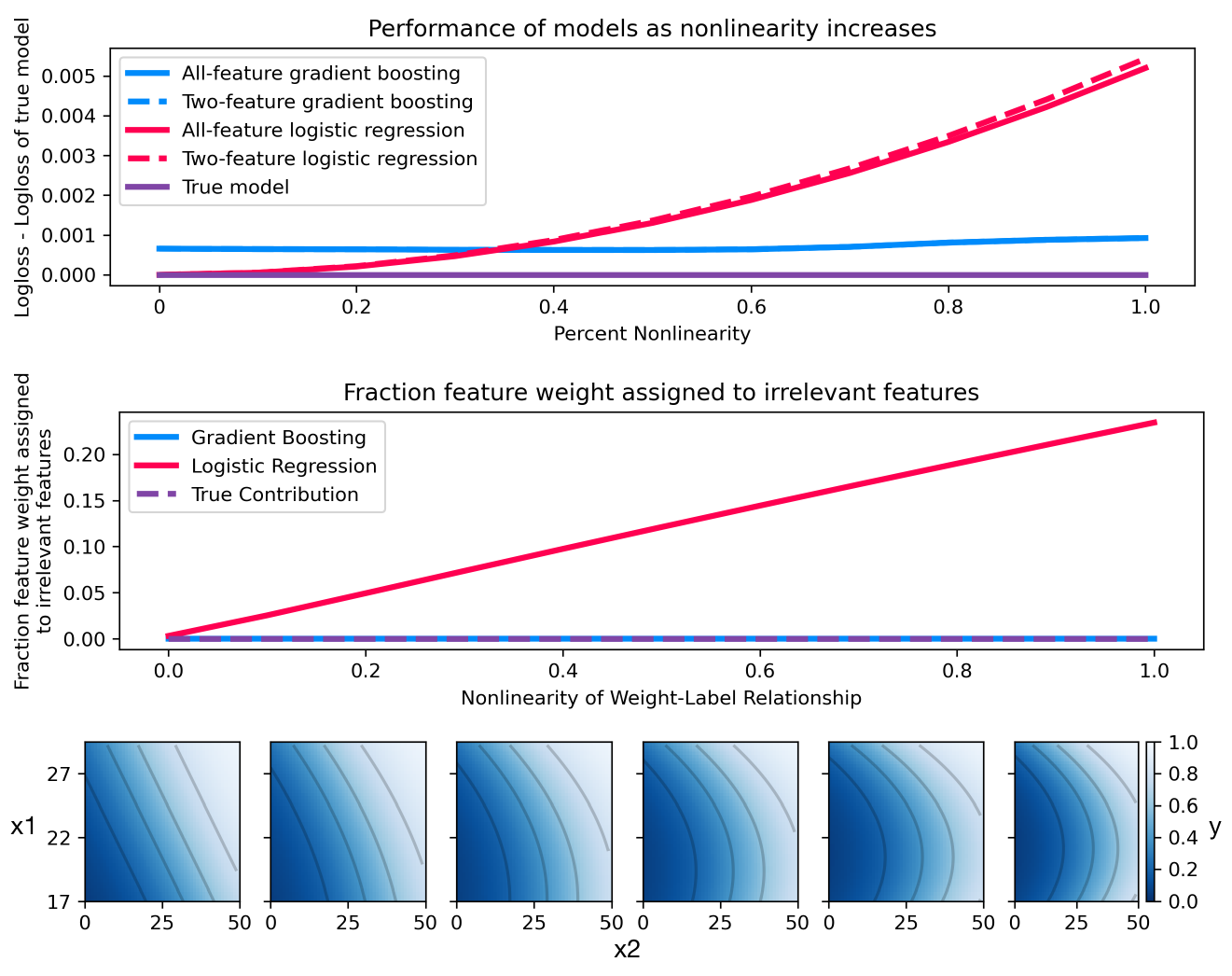}
 
\caption{ Demonstrates how the predictive performance of a logistic regressor drops as non-linearity increases, whereas the XGBoost \citep{Chen2016b} model does not (top); shows how irrelevant feature attribution increases with non-linearity for the linear regressor, but for XGBoost it does not (middle); the relationship between variables in the dataset for these experiments becomes increasingly non-linear. These experiments were close adaptations of those by \citet{Lundberg2020}, original code for which may be found at \url{https://github.com/slundberg/shap}.}
\label{fig:linprobs}
\end{figure}

Here, $\sigma$ is the logistic link function, $q$ is the degree of non-linearity, which is varied between zero (describing a linear relationship) and one (describing a model with a quadratic relationship), $\mathbf{y}$ is the outcome, and $\mathbf{x}_1$ and $\mathbf{x}_2$ are the two predictor variables. The choice of the factors (\textit{e.g.}, $0.388$) and intercepts (\textit{e.g.}, $-0.325$) are arbitrary, and derive from the classic NHANES I dataset \citep{Launer1994, Fang2000} from which the predictors and outcome are drawn. Multiple copies of the dataset are created with different levels of non-linearity between the predictors and the outcome (by increasing $q$ from zero to one). The contour plots in the lowest plot in Figure \ref{fig:linprobs} demonstrate how the relationships between the variables changes with increasing non-linearity. An irrelevant variable is then introduced into the dataset, which does not relate to the outcome (\textit{i.e.}, it is statistically independent). 

Two models were fit to these synthetic data: a linear logistic regressor, and a machine learning algorithm known as \textit{XGBoost} \citep{Chen2016b}. XGBoost is a data-driven algorithm with a flexible/adaptive functional form. The upper plot in Figure \ref{fig:linprobs} shows how the logistic regressor's classification error (cross-entropy loss, or `log loss’) increases as the non-linearity increases. This is expected – the linear model is unable to model the non-linearities in the data. In contrast, the XGBoost model's error remains low, because it is able to adapt to the non-linearity. Notably, when $q$ is close to zero (\textit{i.e.}, the percent non-linearity is low), the linear model actually outperforms the XGBoost model, and has the potential to directly match the data generating process. 

The middle plot shows how the assigned contribution of the irrelevant variable to the model’s outcome changes as the non-linearity increases. Given that this variable is unrelated to the outcome, this contribution should be zero. The assigned contribution is measured in two ways: For the logistic model, the coefficient associated with the variable is used, and for the XGBoost model, we use an explainability technique (discussed further below). Note that, as $q$ increases, the magnitude of the coefficient for the irrelevant variable in the logistic model increases with the non-linearity. This is reflected in the plot as an increase in assigned importance. This is highly problematic for explainability (and clearly also interpretability) - it results in irrelevant variables being assigned predictive importance even when they are not. In contrast, the XGBoost model ignores the irrelevant variable (as one would expect) regardless of the degree of non-linearity, and the associated variable importances (which are derived using a model explainability technique described below) remain near zero.

\subsection*{Explaining Complex Predictive Models - Camels in the Countryside}
In Part 1 we suggested that researchers explore machine learning (ML) methods which facilitate the modeling of complex, non-linear relationships between variables. These techniques are applicable to predictive as well as causal approaches. In the latter case, the ML techniques may be integrated into causal models and perform the role of the arrows in the c-DAGs (there are other ways to integrate ML in causal models, see \textit{e.g.}, \cite{ vanderLaan2018}).

ML models may have many hundreds, thousands, or millions of parameters, and as a result, they are often referred to as `black-boxes’, owing to their opacity both in terms of explanation, as well as interpretation. The benefit of such models is that they provide us with powerful predictive opportunities and, when used correctly, can help us avoid issues relating to functional misspecification.\footnote{By `correctly’ we mean to say that the model balances bias (function inflexibility) and variance (too much flexibility) to achieve an appropriate level of functional complexity to model the relationships in the data.} Furthermore, great strides have been made in developing model explainability techniques, which provide us with a means to interrogate these black-boxes to understand why they make certain predictions. However, explainability does not imply interpretability, and we now describe a famous example which highlights how mitigating functional misspecification does not help us achieve interpretable models. 

One of the principal limitations of purely predictive approaches relates to their inherent structural misspecification (see Part 2). The problem can be demonstrated with an example involving the classification of images of cows and camels, where images of cows frequently feature countryside backgrounds and images of camels tend to feature sandy or desert regions \citep{Arjovsky2020}. A predictive function will not respect the orthogonality and semantics of the animal or background, and the background provides a convenient cue, albeit one which is irrelevant, with which to classify the animal. Hence, a cow in a desert may be wrongly classified as a camel, and a camel with a countryside background may be wrongly classified as a cow. This issue may never become problematic in practice, so long as the function is not exposed to a new distribution of images, where the joint distribution of backgrounds and animals changes. This highlights how predictive models, owing to their structural misspecification, are sensitive to what is known as covariate or distributional shift. This problem is notably resolved when the structure of the problem is appropriately encoded into the model.\footnote{An example of structure being used to independently reason about compositional elements (appearance and motion) in video can be seen in \citet{Vowels2020c}.}

This example concerning issues relating to classification of high-dimensional image data may appear somewhat unrelated to the typical data that psychologists are concerned with, but actually the problem is just as important in the low-dimensional setting. Indeed, predictive models are usually fit by minimizing an error criterion (\textit{e.g.}, mean squared error or binary cross entropy), and there is therefore nothing to restrict these models from leveraging any or all statistical correlations present in the data. The use of predictive model explainability techniques (discussed in more detail below) can be used to help identify whether the model might be leveraging factors which have the potential to be spurious or problematic, and can be used to provide insight, particularly if the included predictors are restricted to those which the researcher \textit{a priori} knows to be causally relevant. For example, a recent paper used model explainability (specifically, random forest variable importance weights) to investigate the predictive generality of theoretically relevant constructs in relationships research \citep{Joel2020}. Unfortunately, the explainability techniques limited to telling us which variables are useful and by how much, rather than anything about the underlying causal structure. Consequentially, predictive models are rarely interpretable without structural constraints.

\subsection*{Explainability Techniques}

The ability to interrogate and explain our predictive models is important, particularly given that the deployment of such models for automated decision making processes have the potential to seriously impact individuals' lives \citep{Hardt2016, Kilbertus2017, locatello2019fairness, cao2019, liu2019, howard2018, rose2010, louizos2017, moyer1, gendershades}. Indeed, the European Union has recently decreed that the use of machine learning algorithms (which includes the use of predictive functions) be undertaken in such a way that any individual affected by an automated decision has the right to an explanation regarding that decision \citep{Aas2019, EU2016}. In the previous section we described the camels in the countryside problem, whereby powerful predictive models with flexible functional form do not respect causal structure in the data. However, complex models (often called \textit{black-boxes}) are more difficult to explain than linear models, and we therefore need explainability techniques to do the explaining for us.

Model explainability is a burgeoning area of machine learning, in which commendable strides have been made in recent years (\textit{e.g.}, \cite{Alaa2019, Wachter2018, Lundberg2020}). The techniques facilitate a form of \textit{meta-modeling}, whereby a simpler, human-interpretable and thereby explainable model is used to represent the more complex, underlying model \citep{Rudin2019}. One popular meta-modeling explainability technique derives from a game theoretic approach to quantifying the contribution of multiple players in a collaborative game; namely, Shapley values \citep{Shapley1953}. Recently, Shapley values have been adapted to yield meaningful explanations of models that correspond well with human intuition \citep{Lundberg2017, Lundberg2019, Lundberg2020, Sundararajan2020, Chen2020b, VowelsSexDes}. Indeed, the Shapley method was used with XGBoost in the experiments demonstrating the problems with linear model interpretability above (Figure \ref{fig:linprobs}). 

The family of Shapley methods provide breakdowns which indicate how much each input variable or feature contributes to a model's prediction for any individual datapoint. They work by conceiving of the variables as the players in a collaborative game, where the aim is to maximize the predictive performance of the model. By iteratively examining how much each variable and each combination of variables contributes towards the performance of the model, the method not only enables one to establish the relative contributions of each variable/variable interaction, but also by how much the model output changes in response to a particular change in each variable/variable interaction.\footnote{Interested readers are directed to \citet{Lundberg2020}.} Such individualized prediction and explanation is particularly important for individualized treatment assignments (for example), and thereby mitigates concerns regarding the use of aggregation in psychology and social science \citep{Bolger2019, Fisher2018}. The methods can be used equally for complex functions (such as neural networks) as well as for simple linear functions. By combining powerful function approximation with explainability techniques, we may be able to achieve accurate forecasts and outcome predictions, while maintaining the capacity to understand what our model is actually doing when it makes a prediction. 

 From a research standpoint, explainability techniques allow researchers to understand, in a purely associational sense, which variables and interactions between variables are important when making a prediction. For example, if one identifies that a variable, previously considered to be important, contributes negligible predictive value then one might investigate whether this variable does or does not fit into a particular theoretical framework. We would therefore argue that researchers should consider a combination of predictive methods with explainability tools as a useful means to contribute new knowledge, particularly during the early and/or exploratory stages of investigation. However, just because a predictive model finds a particular feature (ir-)relevant to making a prediction, does not mean that this degree of association is meaningful outside of the function/model (as with camels in the countryside). Furthermore, an explainability technique represents a form of model in its own right, and the process of modeling a model brings its own difficulties (see \textit{e.g.}, \cite{Rudin2019, Kumar2020}). Indeed, if the explanation model is good at explaining the data in a simple, human-readable form, then the explanation model provides evidence that a simpler, more explainable model was possible to begin with. These difficulties notwithstanding, the explainability techniques provide a valuable means to leverage predictive model for exploratory research. 
 
Finally, and as with functional misspecification and causal misspecification, the distinction between explanation and interpretation stands independently of replicability. However, we acknowledge that it is always important to temper the strength of explanations and/or interpretations according to the sample size and strength of the identified associations or causes.

\subsection*{Part 3 Summary}

In Part 3, we have described how either functional and/or structural misspecification result in uninterpretable models. In such cases, any attempt to interpret the models in spite of these limitations results in conflation and unreliability. The interpretations are conflated because a misspecified model cannot be interpreted causally, and they are unreliable because predictive models can only be explained. This distinction is important because, if a structural misspecfication has occurred (perhaps because we intentionally adopted a predictive/non-causal approach), one can restrict the purview of scientific conclusions to the specific mathematics of the algorithm used for prediction. In other words, powerful function approximation techniques may be able to accurately predict outcomes and have the flexibility to match the functional form of the true data distribution, but they do not necessarily respect or reflect the \textit{causal} structure in the data generating process. 

Does all of this mean that predictive techniques cannot generate understanding? Not entirely. There are many scenarios, particularly during the exploratory stages of a research project, for which researchers may not yet have a strong, empirically supported inductive bias or theory about the data generating process. Rather than testing specific theoretical hypotheses during these early stages, it may be pertinent to ask more general research questions concerning the existence of associations (causal or not). The goal may then be to amass varied evidence (\textit{e.g.}, by using predictive models) to gradually uncover a basis for the development of an increasingly refined theory \citep{Gelman2014, Shrout2018, Oberauer2019, Tong2019}. Furthermore, predictive techniques can be used to evaluate the predictive validity of psychological theory \citep{Yarkoni2017}. Explainability techniques may then be useful in building up an intuition about `what is important' in the phenomenon of interest. However, these techniques are not without their own limitations, and we urge researchers to engage broadly with experts in the practice of these techniques to ensure that (a) their approaches are optimal for their research, and (b) that their interpretations (or explanations) are tempered according to the limitations of their models.

\section*{Recommendations and Conclusion}

The replicability crisis has drawn attention to numerous weaknesses in typical psychology and social science research practice. However, in our view, issues relating to functional misspecification, structural misspecification, and unreliable interpretations have not been sufficiently identified in prior work. These issues are characterized by their shared orthogonality to replication and sample-size. Indeed, facing any of these issues, replication may actually serve to reinforce flawed conclusions, rather than to highlight and rectify them.

\emph{1. We recommend that psychologists and social scientists give more consideration to predictive approaches, particularly during the exploratory stages of a research project.}

The inherent complexity and non-linearity of the typical phenomena of interest to psychologists and social scientists may make the goal of causal inference arbitrarily complex \citep{Meehl1990}. This may partly explain why researchers in psychology and social science are generally discouraged from drawing causal conclusions from observational data, despite them doing so implicitly anyway \citep{Grosz2020, Dowd2011}. Indeed, the use of SEM could be taken as evidence of intention to undertake causal research, as the very structure of the model is an imposition of the researcher's view on the data generating process. The use of an explicit causal graph with opaque predictive interpretations represents a further example of the conflation of predictive and causal approaches. In cases where the models themselves are misspecified both in terms of linear functional form and untestable structural assumptions, the interpretation of such models becomes unreliable.

When researchers wish to model the relationships between variables, either as part of a causal model, or for purposes of prediction alone, then it may be highly advantageous for them to consider techniques common in machine learning, particularly in combination with model explainability techniques. Indeed, \citet{Yarkoni2017} have previously made a similar recommendation. A wide range of powerful predictive modeling techniques exist, including neural networks \citep{goodfellow}, random forests \citep{breiman2001}, gradient boosting machines \citep{Chen2016b}. Such powerful function approximation techniques can be used to leverage as many associations present in the data sample as possible. In the case of predictive modeling, a consideration for the causal structure of the data is possible but not necessary. Incorporating causal inductive bias may aid in generalization, but it is not strictly necessary to achieve good predictive performance. Unfortunately, the use of techniques with potentially data-adaptive, flexible functional form is extremely rare in psychology and social science, where the use of models with restrictive linear functional form is ubiquitous \citep{Yarkoni2017, Blanca2018}.

\emph{2. We recommend that psychologists and social scientists seek collaboration with statisticians and machine learning engineers/researchers, whose principal focus is to understand, practice, and develop predictive and causal modeling techniques.} 

 Given that there exist dedicated fields for the development of modeling approaches (\textit{e.g.}, statistics, machine learning, causal inference), it is perhaps unrealistic to expect an applied researcher in psychology or social science, to have expertise in the practice of predictive and explanatory or causal modeling, particularly when the mathematical knowledge required to understand these techniques is both significant and rare in these fields \citep{dynamic}. Furthermore, new methods are continually being developed and updated.

We therefore recommend that they seek collaboration with experts in the practice of their chosen analytical approach.\footnote{Note that this recommendation has been made by researchers previously in various contexts (\textit{e.g.}, \cite{Lakens2016}).} This recommendation does not abdicate researchers from their responsibility to attain a basic understanding and awareness for these problems, even if they are not the ones ultimately responsible for writing the code or running the analysis. By introducing researchers to these fundamental considerations for research, along with some basic background theory, we hope that this article helps researchers to identify potential areas of weakness in their research process, such that they can seek appropriate collaboration and guidance.

\emph{3. We recommend researchers be transparent about whether they are adopting a predictive or causal approach and to qualify their interpretations.}

Meta-researchers have previously highlighted how psychologists and social scientists tend to mix causal and predictive language \citep{Grosz2020}. Similarly, meta-researchers have drawn attention to how it is common for psychology and social science researchers to use associational/predictive techniques to test otherwise causal hypotheses \citep{Shmueli2010, Yarkoni2017, glymour, Hernan2018, Grosz2020}. This leads to an awkward conflation of causal modeling with predictive interpretations, resulting in ambiguity and a lack of clarity regarding intentions and assumptions. 

We have also discussed how unreliable interpretations may stem from functional and structural misspecification, and how these issues may be common in the fields of psychology and social science. We encourage researchers to ask themselves what an interpretation of an effect size or parameter derived using a naive (\textit{i.e.}, misspecified) model actually means: Is it actually an explanation for how much the output of the \textit{model} changes with respect to a change in the input; or is it being interpreted causally (\textit{e.g.}, this childhood intervention increased well-being by $\theta$-amount)? In either case, researchers need to be transparent and clearly articulate whether they are adopting a predictive or causal approach. Each approach is associated with assumptions and limitations which need to be clearly stated in order to contextualize any explanations or interpretations which are made. Predictive model explainability tools have their own limitations and may actually contradict the results of causal investigations: While the inclusion of a mediator in a regression can completely block a causal path reducing the estimated effect to zero, a strong effect might be indicated by an explanation of a predictive model. 

Similarly to \citet{Grosz2020}, we therefore recommend that researchers clearly state their approach as well as its associated assumptions and limitations, and moderate their explanations, interpretations, and conclusions accordingly.

\emph{4. We recommend that researchers distill their research questions and hypotheses.}
Researchers should distill and simplify causal questions so that they are both minimal and sufficient. For example, in our discussion of causal inference, we chose to identify a single causal effect, and for this it was sufficient to identify the minimal backdoor adjustment set necessary to render this causal effect identifiable. As such, a full graph did not need to be specified, even though it may need to be considered in order to find the backdoor adjustment variables. Researchers should temper their eagerness to specify and interpret large (causal) SEM graphs which are almost invariably accepted as valid \textit{a priori} \citep{vanderweele2020, Ropovik2015}. Here, we appreciate Dawid's \citep{Dawid2008} reference to Bourdieu who warns of "sliding from the model of reality to the reality of the model" \citep{Bourdieu1977}. \citet{vanderLaan2011} also recommend a ``targeted'' approach. More generally, by distilling our research questions and hypotheses, we may be able to increase the chance that our modeling attempts are successful and potentially reduce some of the burden on sample size.

 While we have focused on the fields of psychology and social science, we feel the highlighted issues are relevant to all empirical human sciences fields. There is little doubt that the lack of understanding about the assumptions, limitations, and pitfalls associated with predictive and explanatory modeling represent fundamental issues. Every research question and hypothesis may present its own unique challenges, and it is only through an awareness and understanding of varied statistical methods for predictive and causal modeling, that researchers will have the tools with which to appropriately address them.


\bibliography{NN.bib}
\bibliographystyle{iclr2021_conference}

\end{document}